\documentclass[sigconf]{acmart}

\copyrightyear{2026}
\acmYear{2026}
\setcopyright{cc}
\setcctype{by}
\acmConference[IDE '26]{3rd International Workshop on Integrated Development Environments }{April 12--18, 2026}{Rio de Janeiro, Brazil}
\acmBooktitle{3rd International Workshop on Integrated Development Environments (IDE '26), April 12--18, 2026, Rio de Janeiro, Brazil}
\acmPrice{}
\acmDOI{10.1145/3786151.3788598}
\acmISBN{979-8-4007-2384-1/2026/04}

\usepackage{enumitem}
\usepackage{xspace}
\title[MigMate: A VS Code Extension for Python Library Migration]{MigMate: A VS Code Extension for LLM-based Library Migration of Python Projects}

\author{Matthias Kebede}
\affiliation{
 \institution{New York University Abu Dhabi}
  \city{Abu Dhabi}
  \country{United Arab Emirates}
}
\email{msk9862@nyu.edu}
\orcid{0009-0000-3741-4133}

\author{May Mahmoud}
\affiliation{
 \institution{New York University Abu Dhabi}
  \city{Abu Dhabi}
  \country{United Arab Emirates}
}
\email{m.mahmoud@nyu.edu}

\author{Mohayeminul Islam}
\affiliation{
 \institution{University of Alberta}
  \city{Edmonton}
  \country{Canada}
}
\email{mohayemin@ualberta.ca}

\author{Sarah Nadi}
\affiliation{
 \institution{New York University Abu Dhabi}
  \city{Abu Dhabi}
  \country{United Arab Emirates}
}
\email{sarah.nadi@nyu.edu}

\renewcommand{\shortauthors}{Matthias Kebede, May Mahmoud, Mohayeminul Islam, and Sarah Nadi}

\begin{document}

% article.
\begin{abstract}
Modern software relies heavily on third-party software libraries to streamline the development process. The act of switching one library for a similar counterpart, called library migration, naturally occurs as libraries become outdated or unsuitable for the project. Manually migrating from one library to another is a time-consuming task. 
Our previous research developed MigrateLib, a command-line LLM-based migration tool that can automate the complete migration process.
In this paper, we present our open-source VS Code IDE plugin, MigMate, that builds on MigrateLib by integrating the automated migration process into the developer’s existing development environment.
MigMate provides an interactive experience, allowing developers to view and confirm changes before they are applied.
A preliminary user study shows that plugin usage consistently reduces the time taken to complete a library migration task, and it scores highly on the System Usability Scale.
\end{abstract}

%%
%% The code below is generated by the tool at http://dl.acm.org/ccs.cfm.
%% Please copy and paste the code instead of the example below.
%%



\maketitle

\section{Introduction}
Modern software relies on third-party libraries to streamline development. Libraries provide reusable code and abstraction, but also introduce new maintenance challenges~\cite{trivial-packages}. Managing libraries can mean updating a version or performing a full \textit{library migration}, where one library is replaced by another. Library migrations are difficult and time-consuming when done manually~\cite{do-developers-update}, as developers must learn the Application Programming Interfaces (APIs) of both libraries and then perform code transformations across their codebase. Although some tools recommend replacement libraries or map APIs between the two libraries~\cite{migration-advisor, libcomp, automatic-discovery, deep-diving}, they still leave significant manual effort for the developer.

Accordingly, both ourselves~\cite{py-llm-migration} and other researchers~\cite{llm-migration-results} have turned to investigating whether Large Language Models (LLMs) can be used to fully automate the library migration process, including both API mapping and code transformation.
Our initial empirical study showed that LLMs can correctly perform a high portion of migrations in benchmark evaluations~\cite{py-llm-migration}. Based on the insights gained from the empirical study, we developed an LLM-based end-to-end Command Line Interface (CLI) for library migrations, MigrateLib,  and evaluated it on additional repositories~\cite{pymigtool}.

While standalone tools can be quite powerful, they add friction to the development process by forcing developers out of their workflow~\cite{visualize-info-editors}. Ideally, such tools should be integrated into an Integrated Development Environment (IDE) to provide convenience and improve usability. To enable more streamlined usage of MigrateLib, in this paper, we develop a Visual Studio Code (VS Code) plugin called MigMate. Under the hood, MigMate uses the core functionality of MigrateLib, but adds an interactive UI that integrates into the developer workflow. It allows developers to view any suggested modifications to their code and approve or reject individual changes, giving the developer more control over the migration process and building trust in the tool. In doing so, our work contributes a human-in-the-loop approach to automated library migration that addresses key usability challenges.

We open-source MigMate at \url{https://github.com/sanadlab/MigMate}.
We also provide a video demonstrating how MigMate works at \url{https://www.youtube.com/watch?v=LHEmUFFz8_o}.

\section{Background and Related Work}
\textit{Library migration} is a process where a developer replaces a \textit{source library} with a \textit{target library} that provides similar functionality, without changing the behavior of the project. 
This requires updating the dependencies, updating the API usage in the code, and verifying behavior. Migrations typically happen between \textit{analogous library pairs}, such as \texttt{requests}~\cite{requests-lib} and \texttt{httpx}~\cite{httpx-lib}, which overlap in functionality but differ in syntax and configuration~\cite{migration-advisor, libcomp}. 

Researchers have developed tools to recommend or compare analogous libraries. MigrationAdvisor~\cite{migration-advisor} provides Java recommendations via a web application, while LibComp~\cite{libcomp} integrates suggestions into IntelliJ IDEA. 
These tools lower the burden of selecting a target library but focus mainly on recommendations, not code transformation. API mappings can also be discovered by mining repositories~\cite{automatic-discovery} or analyzing documentation~\cite{deep-diving}.
Other approaches like SOAR~\cite{soar} use synthesis and error messages for automated migration, while recent work explores applications of LLMs for library migration~\cite{llm-migration-results, py-llm-migration}.

In our own previous work~\cite{py-llm-migration}, we used Llama~3.1~(70B), GPT-4o~mini, and GPT-4o on a subset of the code changes found in PyMigBench~\cite{pymigbench}, a benchmark of real-world Python library migrations mined from open-source repositories. This empirical study compared LLM-generated migrations against developer implementations to assess correctness.
A code change is considered correct if it exactly matches the developer's change or is manually identified as a valid alternative.
A migration is considered partially correct if only some developer changes were matched, or fully correct if all changes were matched without making any additional refactoring changes.
The results showed that GPT-4o achieved the highest accuracy, with 94\% of migrations containing at least one correct code change and 57\% of migrations being fully correct.
GPT-4o~mini performed similarly with 93\% of migrations partially correct and 49\% fully correct, while Llama~3.1 trailed behind with 51\% of migrations partially correct and 26\% fully correct.
This evaluation shows that LLMs are already capable of handling complex migrations, with room for improvement.

Using the insights from our empirical study, we designed a CLI tool, MigrateLib, that combines LLMs with pre- and post-processing steps (see Sec. \ref{sec:cli-overview}) to improve migration correctness~\cite{pymigtool}. We find that MigrateLib can migrate 32\% of the migrations with complete correctness. Of the remaining migrations, only 14\% of the migration-related changes are, on average, left for developers to fix. This need for human intervention motivated us to build an IDE plugin to allow for more seamless interaction.

Beyond technical methods of library migration, designing usable developer tools is crucial for integrating the automated process into the IDE. Research shows that developers need encouragement to trust in automated refactoring tools, with clear previews of changes before applying them~\cite{design-refactor-tools}. Visualization studies further suggest that inline displays with optional panels are preferred~\cite{visualize-info-editors}. We adopt these insights in our design of MigMate by combining automated support with human oversight to improve adoption and trust.

\section{MigMate Design}
\textbf{Workflow and UI}. We use guidelines from research on plugin usability~\cite{visualize-info-editors, design-refactor-tools} and VS Code's documentation~\cite{vs-activation, vs-ux-guidelines} to design MigMate to be intuitive and efficient. We specifically design these features:

\begin{itemize}[leftmargin=*]

     \item \textbf{Context-Aware Activation:} MigMate loads lazily using VS Code Activation Events~\cite{vs-activation}. It activates when a workspace contains Python source files or a recognized dependency file (e.g., \textit{requirements.txt}, \textit{pyproject.toml}). This approach prevents unnecessary resource usage when the plugin is not explicitly needed.

    \item \textbf{Plugin Configuration:} MigMate provides flexible configuration options through the VS Code \textit{Settings} interface, allowing developers to adjust plugin behavior to suit their workflows.
    
    \item \textbf{Seamless Migration Trigger:} Developers can start a migration directly from the dependency file with a hover or context menu. These triggers are easily accessible and keep the workflow entirely within the IDE, mitigating the burden of context-switching~\cite{visualize-info-editors}.

    \item \textbf{Guided Library Selection:} A Quick Pick~\cite{vs-ux-guidelines} menu lists source libraries and subsequently prompts the developer to input the name of a target library. Quick Picks are a perfect choice for this step since they work well to facilitate short multi-step inputs.

    \item \textbf{Automated Migration Execution:} MigMate performs the migration, runs the project's test suite, and generates a final preview. It also shows a progress bar during the migration and warns the user in the event of migration errors or test failures.

    \item \textbf{Interactive Migration Preview:} Developers can review changes before applying them selectively. This step addresses known LLM issues such as unrelated edits~\cite{py-llm-migration, llm-driven}, increasing the developer's trust in and control over the tool.

    \item \textbf{Test Results View:} 
    A Webview~\cite{vs-ux-guidelines} displays a summary of migration test results. In the event that one or more tests fail after migration, a warning notification will prompt the user to open the Webview, helping developers investigate failures and identify the root cause.

\end{itemize}

\textbf{CLI Overview}. MigMate uses MigrateLib as the backend, so we first briefly discuss it and its use in our plugin.
\label{sec:cli-overview}

MigrateLib~\cite{pymigtool} takes the names of the desired source and target library as its primary arguments, then proceeds through multiple iterative rounds. The initial \textit{premig} round simply establishes a baseline for the project by running the existing test suite. The \textit{llmmig} round sends the relevant code to an LLM and retrieves migrated content to be applied to the code. Subsequent rounds focus on refining the migration and ensuring its correctness.

MigrateLib leverages test results to determine the status of the migration. 
After the LLM migration, MigrateLib compares the test results before and after migration.
If the results are the same, it considers the migration to be correct and does not proceed any further. 
Otherwise, MigrateLib runs two post-processing rounds trying to correct the migrations (specifically re-including code where the LLM says that the rest of the code stays the same and adding the async keyword to function definitions that use asynchronous libraries), and similarly runs the tests to verify the migration.
MigrateLib preserves the test reports of each round for the user to review.

\textbf{Integration with MigrateLib}. MigMate acts as an intermediary between VS Code and the underlying MigrateLib CLI tool~\cite{pymigtool}. When starting a migration, MigMate spawns a child process to run MigrateLib in the background. The source and target libraries are passed to MigrateLib's initiating command, along with certain configuration options as additional arguments. MigMate then parses the migration output for use in the Migration Preview and Test Results views.

    \begin{figure} %*
        \centering
        % \resizebox{0.75\textwidth}{!}{
        \includegraphics[width=0.5\textwidth]{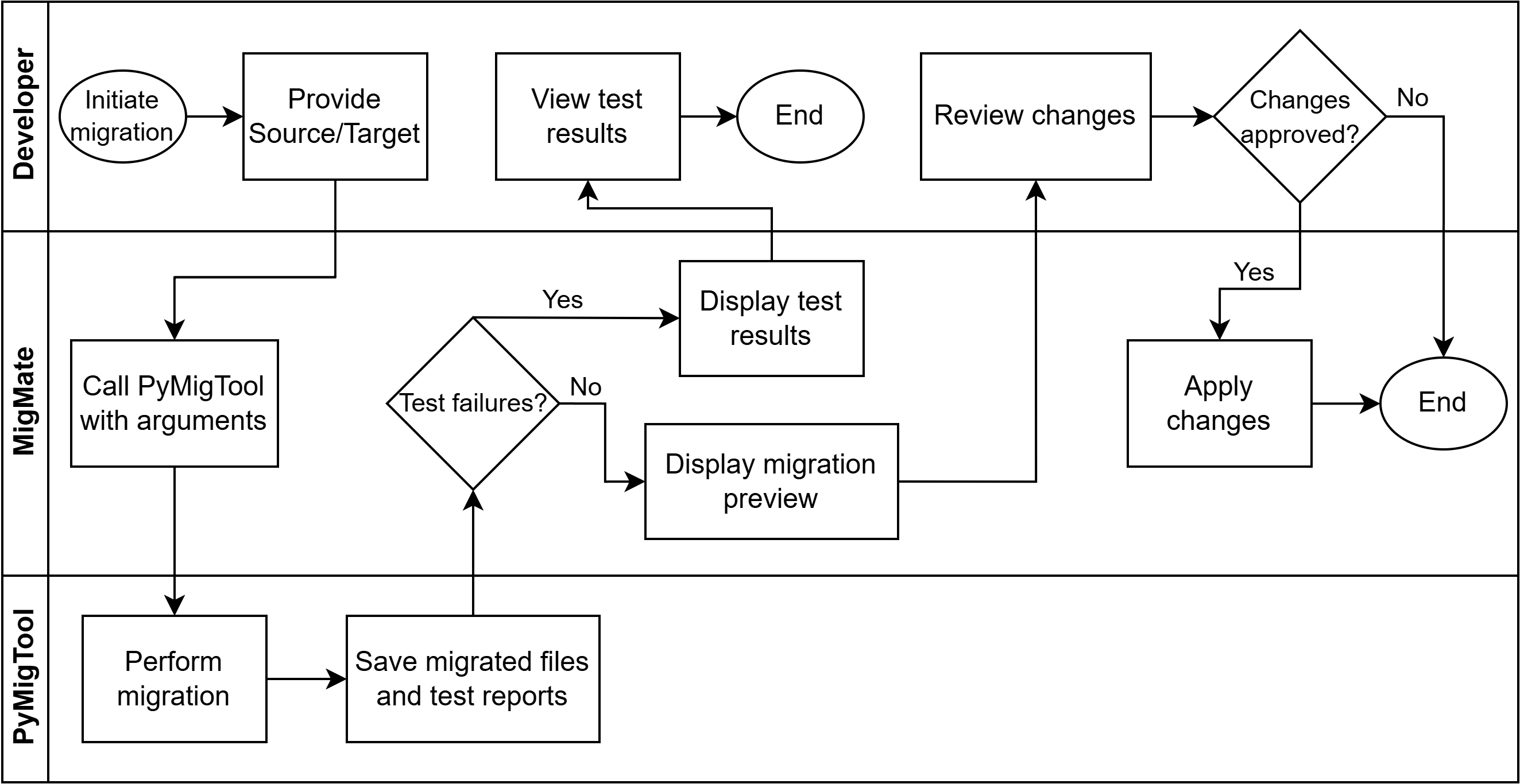}
        % }
        \caption{MigMate Workflow}
        \label{fig:workflow}
        \Description{Workflow diagram showing plugin process}
    \end{figure} %*

\section{MigMate Workflow}
In order to illustrate the expected developer workflow, we present a sample migration from \texttt{requests} to \texttt{httpx} using MigMate.
The following are the high-level steps involved, which are explained in detail in the relevant subsections and shown in Fig.~\ref{fig:workflow}.

\begin{enumerate}[leftmargin=*]
    \item \textbf{Activate Plugin:} The developer opens a Python project in VS Code. MigMate automatically activates after detecting the appropriate source files.
    \item \textbf{Initiate a Migration:} The developer then opens the dependency file (\textit{requirements.txt}) and hovers over \textit{requests}, then selects \verb|Migrate requests| (Fig.~\ref{fig:trigger-hover}).
    \item \textbf{Select Libraries:} \textit{requests} becomes the source library for the migration. The developer enters \textit{httpx} into the provided text input to set the target library.
    \item \textbf{Run Migration:} MigMate calls MigrateLib using the selected libraries and current configuration. A progress bar is displayed in the notification area.
    \item \textbf{View Test Results:} MigMate parses MigrateLib's test report files and provides a summary comparing pre-migration test results to post-migration results (Fig.~\ref{fig:test-results}).
    \item \textbf{Review Proposed Changes:} MigMate opens a Migration Preview window (Fig.~\ref{fig:mig-preview}). The developer selects the desired changes to be applied, then closes the preview.
    \item \textbf{Apply Changes:} MigMate applies the approved changes to the source files. 
\end{enumerate}

\subsection{Initiating Migration}
\begin{figure}[t!]
        \centering \includegraphics[width=0.35\textwidth]{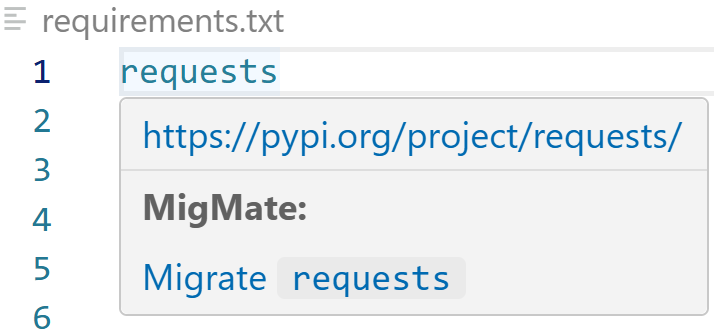}
        \caption{Hover Trigger}
        \label{fig:trigger-hover}
        \Description{VS Code hover menu with a migration button}
    \end{figure}

MigMate provides two main ways to initiate a migration. Given that the dependency file contains the Python libraries used in the project, it serves as a natural starting point for the migration process.
The first approach involves right-clicking within the dependency file to bring up a context menu that provides a migration command. 
Alternatively, the user can move the cursor over the name of a library in the dependency file and wait for a hover menu to appear (Fig.~\ref{fig:trigger-hover}).

\begin{figure}[t!]
        \centering
        \includegraphics[width=0.4\textwidth]{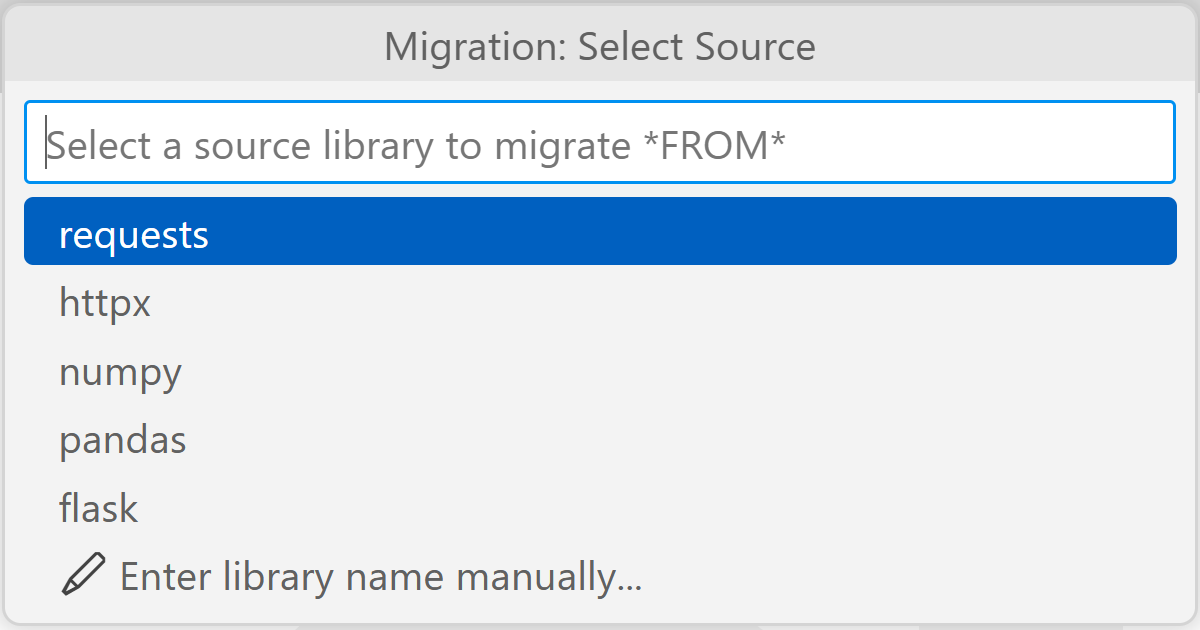}
        \caption{Source Library Selection}
        \label{fig:quick-pick}
        \Description{VS Code Quick Pick menu showing source library options}
    \end{figure}

Regardless of the trigger used, MigMate will present a Quick Pick menu (Fig.~\ref{fig:quick-pick}) to the user where they can input the source and target library names. When using the hover trigger, the source library will automatically be set as the library used to initiate it. Developers can also initiate a migration directly through the Command Palette. Selecting the 'MigMate: Migrate a Library' command will open a Quick Pick just as when using the context menu trigger.

\subsection{Test Results}
\begin{figure}[t!]
    \centering
    \includegraphics[width=0.5\textwidth]{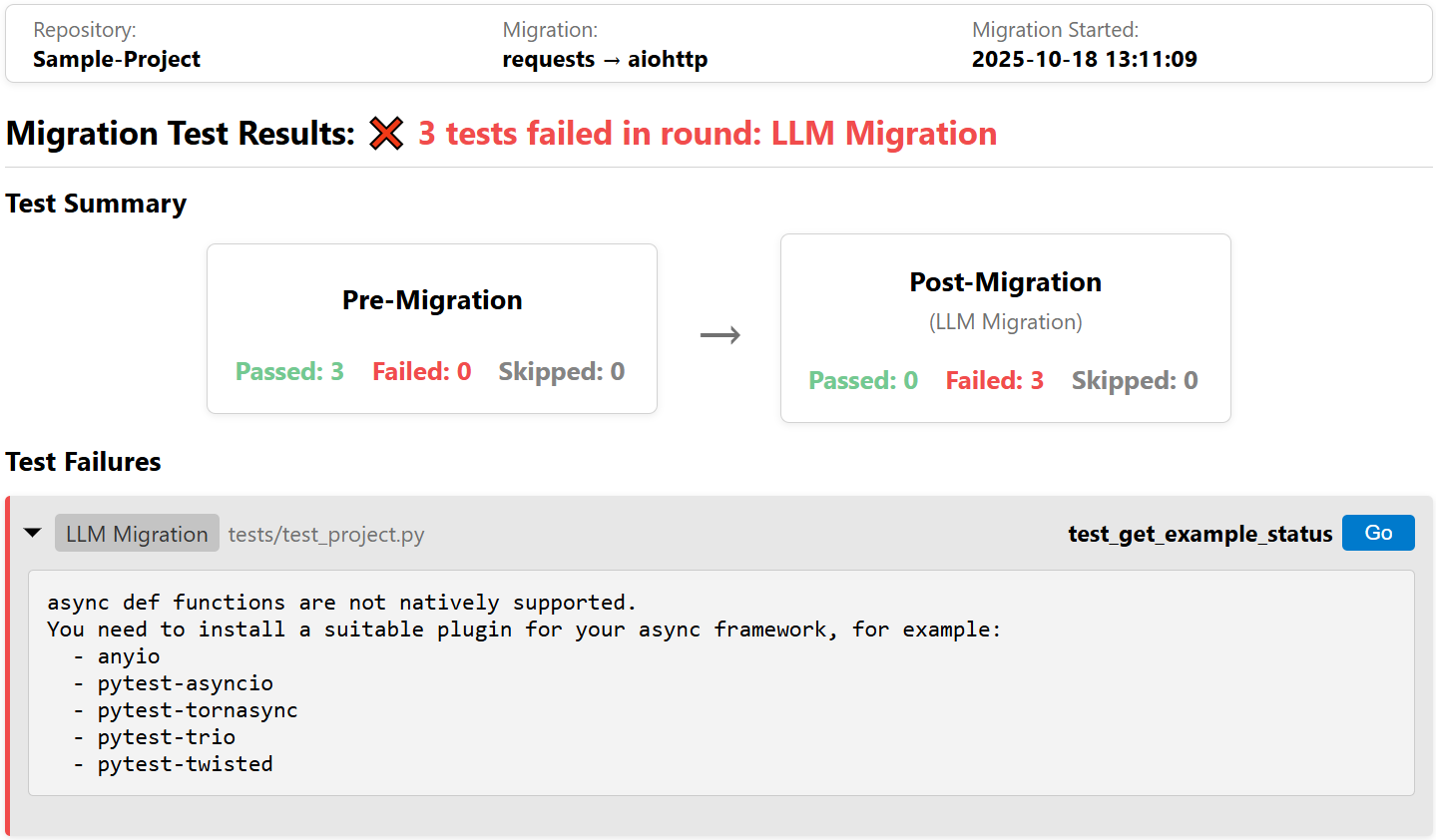}
    \caption{Migration Test Results}
    \label{fig:test-results}
    \Description{VS Code Webview showing test results of a migration}
\end{figure}

MigMate also offers a test result viewer, which the user is prompted to open in the event that any tests fail at the end of the migration process.
This feature is implemented as a Webview, an embedded browser panel controlled by the plugin.
The viewer reads the test report files produced by MigrateLib and renders the data as formatted HTML (Fig.~\ref{fig:test-results}). This allows developers to view a summary of pre- and post-migration test results, the specific error messages of each test, and the contents of the log file created during migration. Clicking the 'Go' button next to each test navigates to the relevant test file.

\subsection{Migration Preview}
A key feature of MigMate is that it requires explicit user approval before applying any migration changes. To generate the preview, MigMate compares the unmodified workspace files with the migrated copies saved by MigrateLib. Two preview styles are available, each offering a convenient interface for reviewing and managing changes across multiple files. 

The first style leverages VS Code's built-in Refactor Preview window, which lists all suggested modifications by file. Developers can selectively enable or disable individual changes using checkboxes, providing fine-grained control over which are accepted. Upon confirmation, MigMate applies all approved modifications in a single bulk edit. Unselected changes are safely discarded without impacting the code.

\begin{figure*}[t!]
    \centering
    \resizebox{0.7\textwidth}{!}{
    \includegraphics{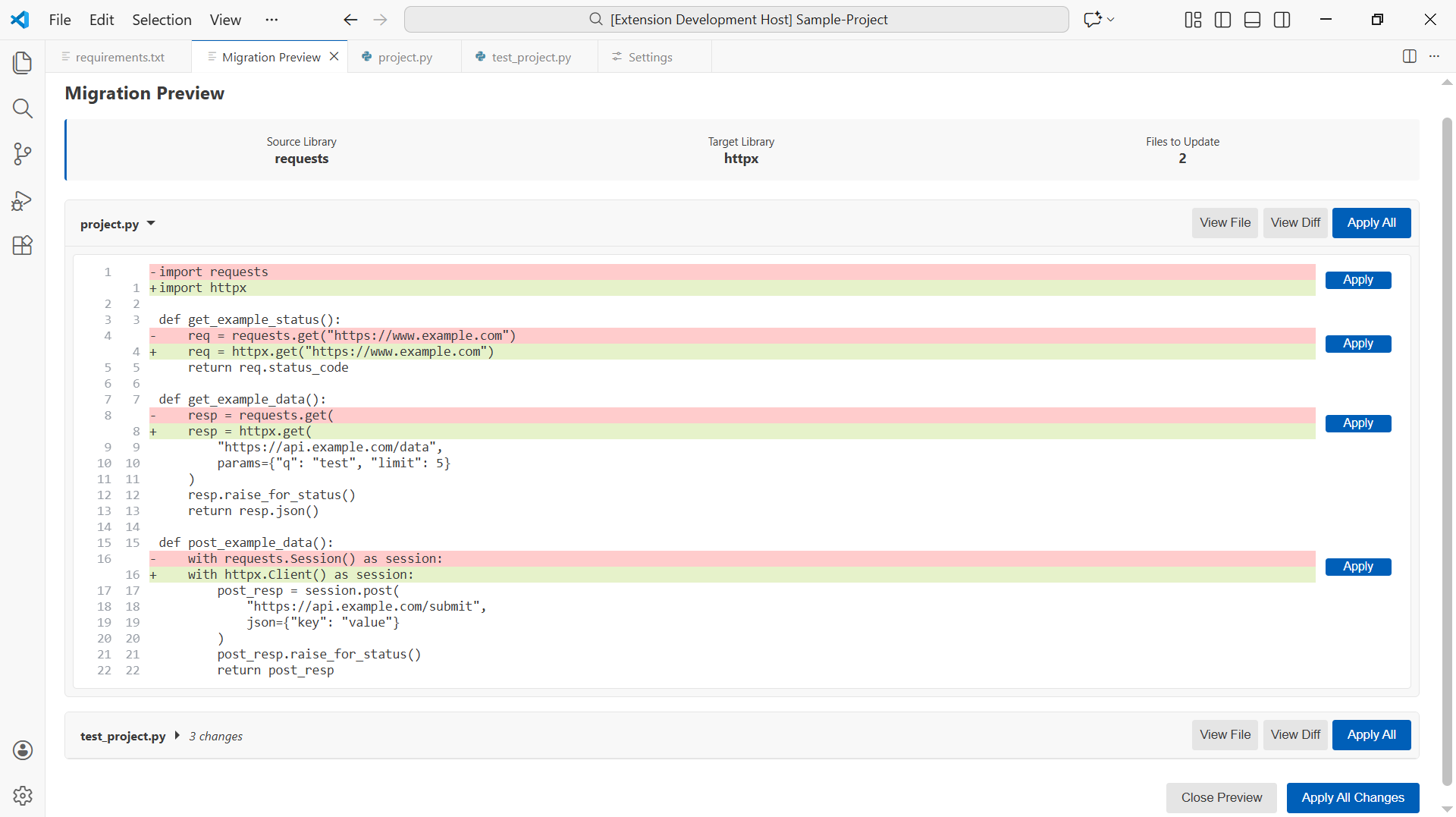}
    }
    \caption{Migration Preview (Webview) for \textit{requests} to \textit{httpx}}
    \label{fig:mig-preview}
    \Description{VS Code Webview showing an interactive diff preview of migration changes}
\end{figure*}

Alternatively, a custom Webview interface (Fig.~\ref{fig:mig-preview}) displays a collapsible list of files and supports the incremental application of edits. With this style, developers can apply individual changes, all changes within a single file, or the entire migration at once. If the migration is accepted as a whole, the Webview will close automatically. Otherwise, the user manually exits by selecting the `Close Preview' button once they are satisfied with the migration changes that they have already applied.

Both preview styles enable detailed selection of edits but differ in terms of when those edits are performed. The Refactor Preview aggregates approved changes into one confirmed update, whereas the Webview applies them incrementally as the developer proceeds.
In either case, MigMate ensures that every modification is explicitly approved, preventing unintended modifications by the LLM and reinforcing developer trust in the migration process.

\begin{figure}[t!]
    \centering
    \resizebox{0.40\textwidth}{!}{
        \includegraphics{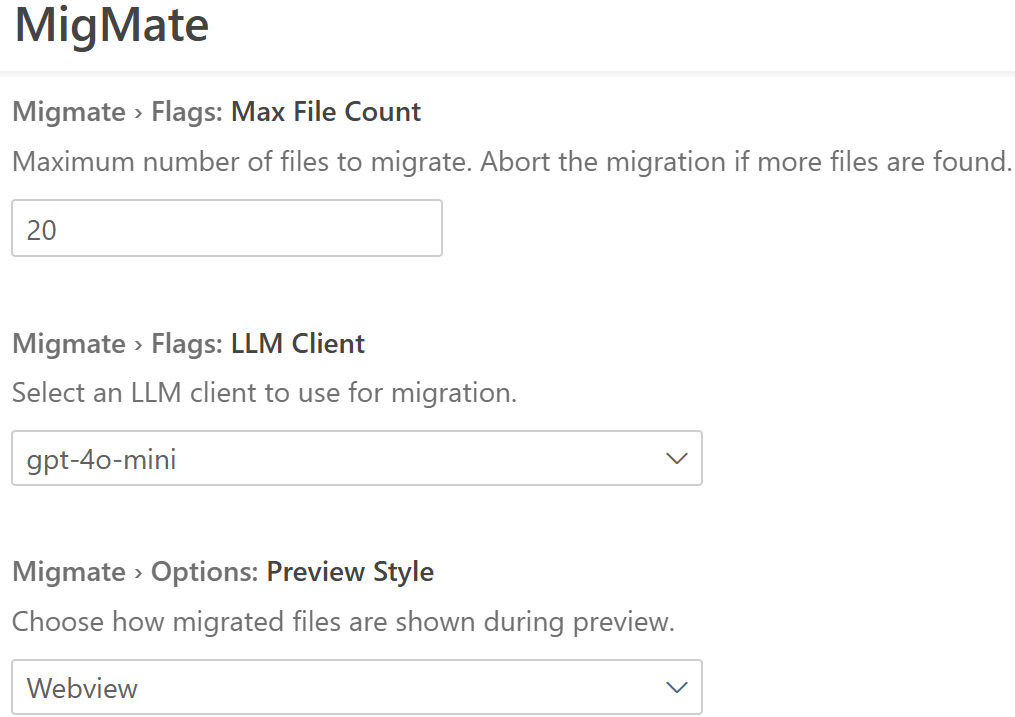}
    }
    \caption{Plugin Configuration}
    \label{fig:config}
    \Description{VS Code settings menu showing plugin configuration options}
\end{figure}

\subsection{Configuration} 
Before performing a migration, developers may wish to customize how MigMate operates. Configuration options are accessible through the VS Code \textit{Settings} interface (Fig.~\ref{fig:config}) or by editing a project-level \textit{settings.json} file. The available parameters are organized into two categories based on their purpose:

\textbf{Migration Flags:} These settings mirror the command-line arguments of MigrateLib and control how migrations are executed. One notable item is the \verb|--llm| flag, which allows the user to pass the name of the LLM model that they wish to use. MigrateLib currently supports OpenAI models, with GPT-4o~mini being the default model, and checks for an \verb|OPENAI_API_KEY| environment variable.

\textbf{Extension Options:} These configurations adjust plugin-specific usability features. Developers can choose between the Webview or Refactor Preview as their preferred migration preview method and toggle whether a preview is shown in the event of post-migration test failures.

\section{Preliminary Evaluation}
To evaluate the usability of MigMate, we conduct a preliminary small-scale user study with undergraduate students at NYU Abu Dhabi.
We ask participants to perform library migration tasks both using MigMate and manually while collecting their feedback.

\subsection{Study Design}
We provide participants with a Python project that makes use of two libraries, \textit{requests} and \textit{tablib}, which can be replaced with \textit{httpx} and \textit{pandas}, respectively. We refer to these as Pair A (\textit{requests → httpx}) and Pair B (\textit{tablib → pandas}). After completing a short warm up task to familiarize them with the project's code, we ask participants to perform library migration tasks as follows:

\begin{itemize}
    
    \item \textbf{Manual Migration:} Participants perform one migration task without using MigMate. They may access API documentation and other online resources, but cannot use LLMs to assist them. This serves as a baseline measurement.
    
    \item \textbf{Plugin-Assisted Migration:} Participants perform a second different migration task with the help of MigMate. They may additionally make use of the same resources as in the manual migration task.

\end{itemize}

\begin{table}[t!]
    \centering
    \caption{Participant Grouping}
    \label{table:study-groups}
    \begin{tabular}{cc cc c}
        \toprule
        \textbf{Group} & \textbf{1st Task} & \textbf{Lib Pair} & \textbf{2nd Task} & \textbf{Lib Pair} \\
        \midrule
        A1 & Manual & A & Assisted & B \\
        A2 & Assisted & A & Manual & B \\
        B1 & Manual & B & Assisted & A \\
        B2 & Assisted & B & Manual & A \\
        \bottomrule
    \end{tabular}
    
    \Description{Table showing four groups and their corresponding task orders}
\end{table}

\subsubsection{Tasks}For Pair A, the project contains 14 \textit{requests} usages spread across 2 files. Most of these are one-to-one changes that replace individual function calls and imports with their \textit{httpx} counterparts. In addition, there are a few more complicated many-to-one changes due to inherent differences between the two libraries. Importantly, migrating this pair does not require any async usage. For Pair B, there are 22 \textit{tablib} usages across 2 files, including a few one-to-one changes but primarily focusing on a mix of many-to-one and many-to-many changes.

For each task, participants have a maximum of 30 minutes to complete the migration. 
 Before starting the plugin-assisted migration, we verbally instruct the participants on how to use the plugin and show them the README file for MigMate.
  The exact project and task instructions used are available at \url{https://github.com/sanadlab/MigMate-Study-Repo}. 
 
 \subsubsection{Survey} After completing both tasks, we ask participants to fill in a short survey that contains Likert-type questions on perceived usability of MigMate based on the System Usability Scale (SUS)~\cite{sus-usability-scale}, a widely used questionnaire that measures perceived usability. The survey also includes optional feedback questions to gather qualitative data regarding the participants' experience with MigMate.

\subsubsection{Participants Block Assignment} We conduct our study as a within-subjects design \cite{within-subjects}, where all participants are exposed to both migration tasks. We balance the participants across different setups (experimental blocks) of the study to mitigate learning effects, especially since the same project is used across both tasks. Half of the participants perform the manual migration first, while the other half begin with the plugin-assisted migration. We further divide those groups into those that start with Pair A and those that start with Pair B, for a total of four groups. Table \ref{table:study-groups} shows the four different experimental block configurations.

\begin{table}[t!]
    \centering
    \caption{Average Time to Complete Migration (min)}
    \label{table:time-averages}
    \Description{Table showing average time taken to migrate each library pair manually or with the plugin}
    \begin{tabular}{lrr}
         \toprule
         \textbf{Lib Pair} &  \textbf{Manual} & \textbf{Plugin} \\ \hline
         \textit{(A) requests → httpx} & 25:23 & 10:42\\ 
         \textit{(B) tablib → pandas} & 27:51 & 10:48\\ \bottomrule
    \end{tabular}
    
\end{table}

\subsubsection{Collected data}
We collect the time taken for each task, as well as telemetry data to identify usage patterns. Note that we do not measure the correctness of the migrations as our main focus is on understanding the participants' experience in using the plugin.

\subsection{Participant Recruitment}
We recruited nine participants for the study, but one withdrew before completing the session. The remaining eight participants consisted of two first-year students, a third-year student, four fourth-year students, and one recent alumnus. All participants self-rated their Python skills and familiarity with VS Code on a 5-point Likert scale, and we included only those who indicated at least a 3 in both areas.

\subsection{Results}

%\subsubsection{Task Timing Results}
Table \ref{table:time-averages} shows the average time taken to solve each task manually versus using MigMate.
We find that across both tasks, the manual migration took more time to complete than the assisted one.  We can see that the plugin saves 60\% of the time required on average.

Based on the telemetry data, we find that the Hover Trigger was the most frequently used method for initiating a library migration by participants (62.5\%), followed by the Context Menu (33.3\%). This matches our expectations, as the hover trigger reduces friction by eliminating the step of selecting a source library for the current migration. The Command Palette was used only once, suggesting it may be less intuitive for users. We find that four participants used the migration-level changes, three used the file-level changes, while one participant completed the migration by exclusively making individual changes. This result does not indicate the exact reasoning behind these choices, which could be due to confidence or simply personal preference. We need further investigation to better understand if a particular granularity level has advantages/preferences.

% \subsubsection{Survey Results}
Based on the questionnaire, we find that MigMate's mean SUS usability score is 80.9 on the 100-point scale, placing it around the 90th percentile and earning an A-grade \cite{sus-interpretation}. 
Six of the participants also left comments and suggestions in the optional feedback. One recurring request was to implement library recommendations during the target library selection when initiating a migration. Another common point was an appreciation for the preview's clarity, with a few participants suggesting that it should also indicate the confidence in each suggested change.

\subsection{Threats to Validity \& Future Work}
Our current evaluation is a small-scale preliminary evaluation of MigMate.
Specifically, our sample size of 8 participants is rather small and the four experiment blocks do not all have the same number of participants. Specifically, Group B1 has three members while Group A2 has only one due to random assignment and some participants canceling their sessions. Another threat is that the participants had limited exposure to MigMate's features. Since they only used the plugin to complete one migration task, they did not need to interact with the configuration options, and most of the participants never experienced test failures during an assisted migration. While we currently have initial positive results about MigMate's usability, future larger scale evaluations should include migration tasks that are known to result in failing tests and to allow participants to experiment with the configuration options.
Such an extended evaluation can also helps us determine useful features to add to MigMate, as well as how to improve the existing functionality.

\section{Conclusion}
This paper presented MigMate, a VS Code extension that integrates MigrateLib~\cite{pymigtool}, an LLM-based migration tool, to support semi-automated Python library migration. The system combines automated code transformation with interactive review of LLM-generated code changes, allowing developers to selectively apply suggested changes directly in their workspace. By embedding this process into the IDE, MigMate streamlines migration while maintaining developer control over code modifications.
Our preliminary evaluation suggests that developers appreciated having a preview of the exact changes that will happen.
More broadly, this work demonstrates how automation can be applied in ways that preserve transparency and confidence in AI-assisted tools.

\begin{acks}
    We would like to thank Ajay Kumar Jha for providing feedback in the early stages of this study.
\end{acks}

\bibliographystyle{ACM-Reference-Format}
\bibliography{paper}

\end{document}